\documentclass{blois}

\newcommand{\Photo}{}

\begin{document}
\vspace*{4cm}
\title{LEPTOPHILIC NEW PHYSICS AND THE CABIBBO ANGLE ANOMALY}

\author{FIONA KIRK}

\address{
Physik-Institut, Universit\"at Z\"urich, Winterthurerstrasse 190, CH--8057 Z\"urich, Switzerland\\
Paul Scherrer Institut, CH--5232 Villigen PSI, Switzerland}

\maketitle
\abstracts{
The Cabibbo Angle Anomaly, an apparent deficit in first-row CKM unitarity, can be addressed by \emph{leptophilic} Standard Model extensions that generate new contributions to the Fermi constant and affect the determination of the CKM element $V_{ud}$.
We focus on simplified models with this property, including the Standard Model extended by vectorlike leptons, by the singly charged scalar singlet, or by a leptophilic $Z'$ boson.
}

\section{Motivation}
The Cabibbo Angle Anomaly~\cite{Hardy:2020qwl,Seng:2018yzq,Czarnecki:2019iwz} is a $\approx 3\sigma$~\cite{Zyla:2020zbs} deviation from first-row CKM unitarity~\cite{Zyla:2020zbs}
 $$|V_{ud}|^2 + |V_{us}|^2 + |V_{ub}|^2 =0.9985(3)(4)\,,$$
where the first error is the uncertainty from $|V_{ud}|^2$, which is extracted from superallowed $\beta$ decays, and the second error is the uncertainty from $|V_{us}|^2$, that can be obtained from $K$ decays and $\tau$ decays.
The Cabibbo Angle Anomaly is known as such because $V_{ud}$ and $V_{us}$, which are the two numerically relevant first-row CKM elements, can be parametrised by the Cabibbo angle.

In these proceedings we assume the Cabibbo Angle Anomaly to be due to new physics~\cite{Grossman:2019bzp,Kirk:2020wdk}, in particular to lepton flavour universality violation~\cite{Crivellin:2020lzu} induced by \emph{leptophilic} new physics~\footnote{The Cabibbo Angle Anomaly can also be resolved by a modified $Wud$ coupling~\cite{Belfatto:2019swo} or by new tree-level contributions to superallowed $\beta$ decays.~\cite{Crivellin:2021rbf}}. Since besides the deficit in first-row CKM unitarity also a deficit in first-column CKM unitarity has been observed~\cite{Zyla:2020zbs}, and since $V_{ud}$ is the largest first-row (and first-column) CKM element, making it the element with the largest impact on the unitarity relations in question, 
we focus on \emph{leptophilic} new physics affecting the extraction of $V_{ud}$ from superallowed $\beta$ decays.

\begin{figure}
\centering
\begin{minipage}[t!]{.48\textwidth}
\centering
\includegraphics[width=.79\textwidth]{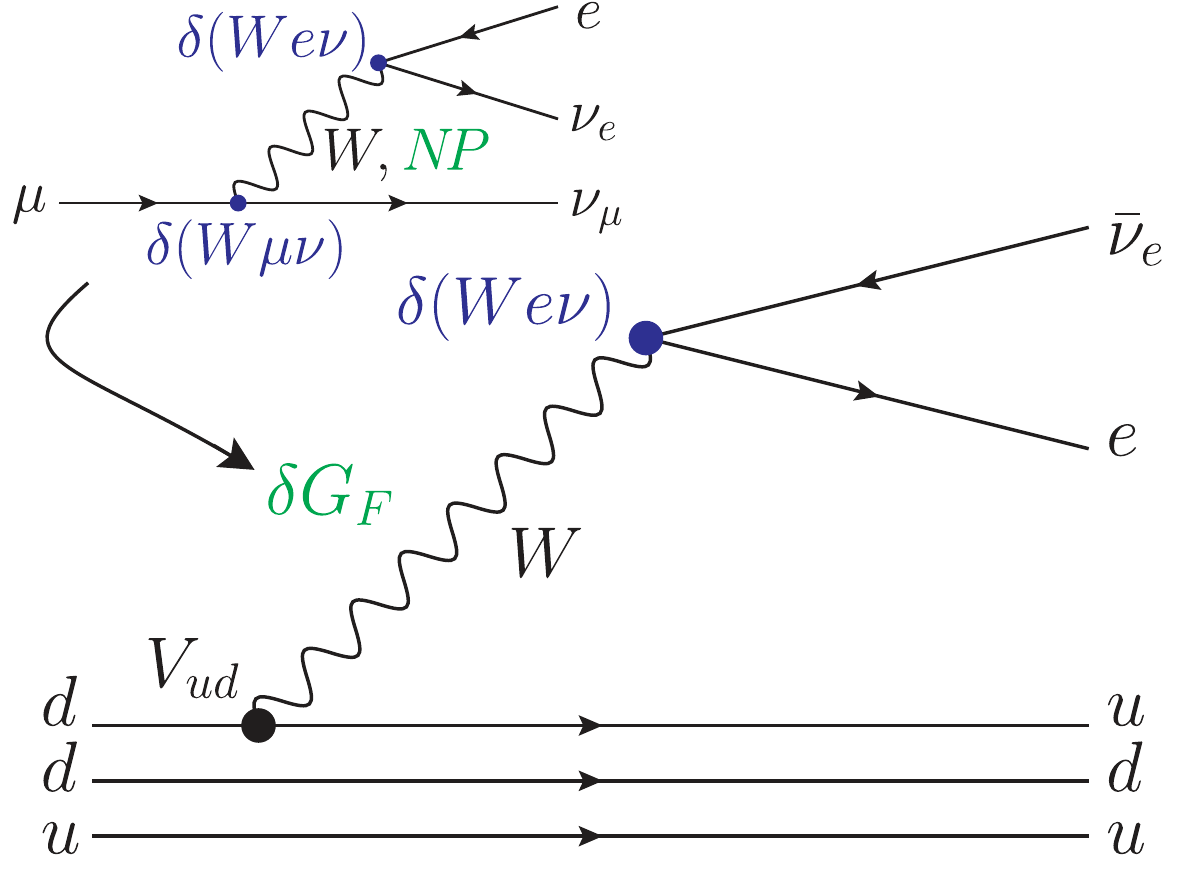}
\caption{Leptophilic resolutions of the Cabibbo Angle Anomaly. New tree-level contributions to muon decay ($NP$) are indicated in green, modified gauge boson couplings to SM leptons are shown in blue. Both $NP$ and $\delta(W\mu\nu)$ affect the determination of $V_{ud}$ from superallowed $\beta$ decays. \label{fig:CAAleptophilic}}
\end{minipage}
\hfill
\begin{minipage}[t!]{.48\textwidth}
\centering
	\includegraphics[width=.74\textwidth]{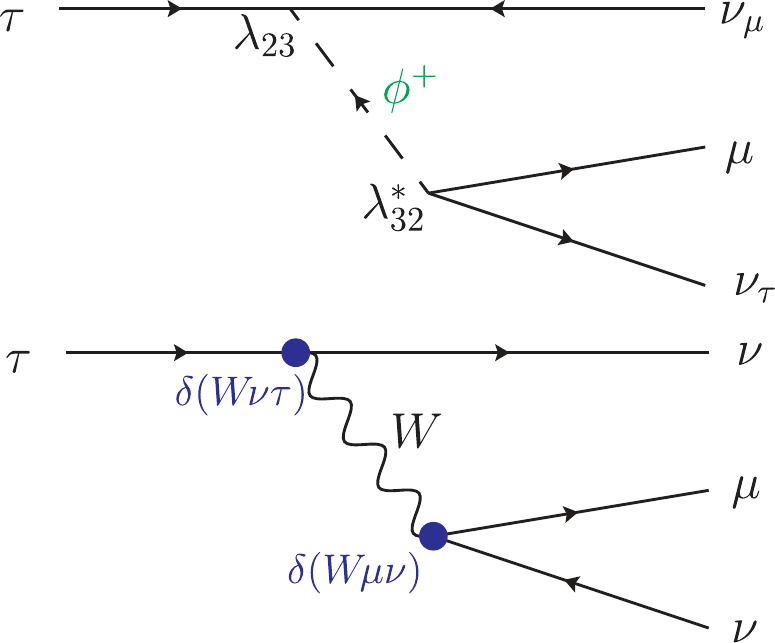}
\caption{Leptophilic new physics inducing lepton flavour universality violation in $\tau$ decays. The first diagram shows a tree-level contribution of a simplified SM extension (here: the singly charged scalar singlet) to $\tau\to \mu\bar\nu\nu$, the second diagram illustrates how modified $W\ell\nu$ couplings can lead to a modification of $\tau\to \mu\bar\nu\nu$. \label{fig:LFUVleptophilic}}
\end{minipage}
\end{figure}

Figure~\ref{fig:CAAleptophilic} illustrates the two ways of resolving the Cabibbo Angle Anomaly with \emph{leptophilic} new physics, both involving new contributions to muon decay, which defines the Fermi constant, $G_F$. As indicated in green, one strategy is to introduce new tree-level contributions to muon decay. The Fermi constant then takes the form $G_F=G_F^{\mathrm{SM}}(1+ \delta(\mu\to e\bar\nu\nu))$, where $G_F^{\mathrm{SM}}$ is the Fermi constant in absence of new physics. Since only $G_F^{\mathrm{SM}}=G_F(1- \delta(\mu\to e\bar\nu\nu))$ enters superallowed $\beta$ decays, 
this allows first-row CKM unitarity to be restored:
\begin{equation}
|V_{ud}|^2(1-\delta(\mu\to e\bar\nu\nu))^2 + |V_{us}|^2 + |V_{ub}|^2 =0.9985(3)(4)\,.
\label{eq:frCKMmod}
\end{equation}
If $V_{us}$ is extracted from $K\to \pi e\nu$, it is also reduced to $V_{us}(1-\delta(\mu\to e\bar\nu\nu))$, however, as $|V_{ud}|^2/|V_{us}|^2\approx 20$, this modification is less crucial than that to $V_{ud}$.

Alternatively, the Cabibbo Angle Anomaly can be resolved by a modified $W\mu\nu$ coupling (shown in blue in Figure~\ref{fig:CAAleptophilic}), which enters superallowed $\beta$ decays, leading to a similar situation as in Eq.~(\ref{eq:frCKMmod}).
The modification of the $We\nu$ coupling does not enter the determination of $V_{ud}$ as a result of a cancellation between the modified Fermi constant and the direct contribution of the modified $We\nu$ coupling to $\beta$ decays. It can affect the extraction of $V_{us}$ from $K\to \pi \mu\nu$, however, its numerical impact on the unitarity relation is very small. 

Figure~\ref{fig:LFUVleptophilic} illustrates how the same kinds of modifications can also address the hints for lepton flavour universality violation suggested by the coupling ratios $g_\tau/g_e$, $g_\mu/g_e$ and $g_\tau/g_e$ fitted from pure leptonic processes by HFLAV~\cite{HFLAV:2016hnz}. Applying the HFLAV results to amplitudes of the type $\ell\to \ell'\bar\nu \nu$, we obtain the amplitude ratios
\begin{small}
\begin{eqnarray}
\frac{\mathcal{A}(\tau  \to \mu  \bar \nu \nu )}{\mathcal{A}(\mu  \to e \bar \nu\nu )} =\frac{g_\tau}{g_e}
= 1.0029(14),\quad
\frac{\mathcal{A}(\tau  \to \mu  \bar \nu\nu )}{\mathcal{A}(\tau  \to e \bar\nu \nu )} =\frac{g_\mu}{g_e}
 = 1.0018(14) ,\quad
\frac{\mathcal{A}(\tau  \to e \bar \nu \nu )}{\mathcal{A}(\mu  \to e\bar \nu \nu )} =\frac{g_\tau}{g_\mu}
 = 1.0010(14)\,,
\label{eq:LFUratios}
\end{eqnarray}
\end{small}
which indicate a slight preference for a new physics effect in $\mathcal{A}(\tau\to\mu\bar\nu\nu)$.

\section{Examples of Leptophilic New Physics}
Let us now focus on simplified \emph{leptophilic} models that can address the Cabibbo Angle Anomaly and the hints for lepton flavour universality violation suggested by the amplitude ratios in Eq.~(\ref{eq:LFUratios}). 

A modification of the $W\mu\nu$ coupling ($\delta (W\mu\nu)$ in Figures~\ref{fig:CAAleptophilic} and \ref{fig:LFUVleptophilic}) can be induced via mixing of the SM $W$ boson with a $W'$ boson that couples to left-handed leptons. This is the case for a $W'$ embedded in a vector triplet with $Y=0$. 
However, a vector triplet with $W-W'$ mixing would also be plagued by $Z-Z'$ mixing, leading to new contributions to $Z\to \ell\ell$, which is strongly constrained by LEP bounds~\cite{ALEPH:2013dgf}.
Alternatively, the $W\mu\nu$ coupling can be modified via mixing of vectorlike leptons with the SM leptons that is induced by electroweak symmetry breaking. 

SM extensions leading to suitable tree-level effects in muon decay (shown in green in Figures~\ref{fig:CAAleptophilic} and \ref{fig:LFUVleptophilic}) are the singly charged scalar singlet~\cite{Crivellin:2020klg}, a leptophilic $Z'$ boson with flavour off-diagonal couplings~\cite{Buras:2021btx}, or the $W'$ boson of a vector triplet.
The scalar triplet would also lead to a new tree-level contribution to muon decay, it would, however, lead to a negative value for $\delta(\mu\to e\bar\nu\nu)$, increasing the tension with first-row CKM unitarity instead of decreasing it.

\begin{figure}
\begin{minipage}[t]{.47\textwidth}
\phantom{ghost}
\includegraphics[width=.9\textwidth]{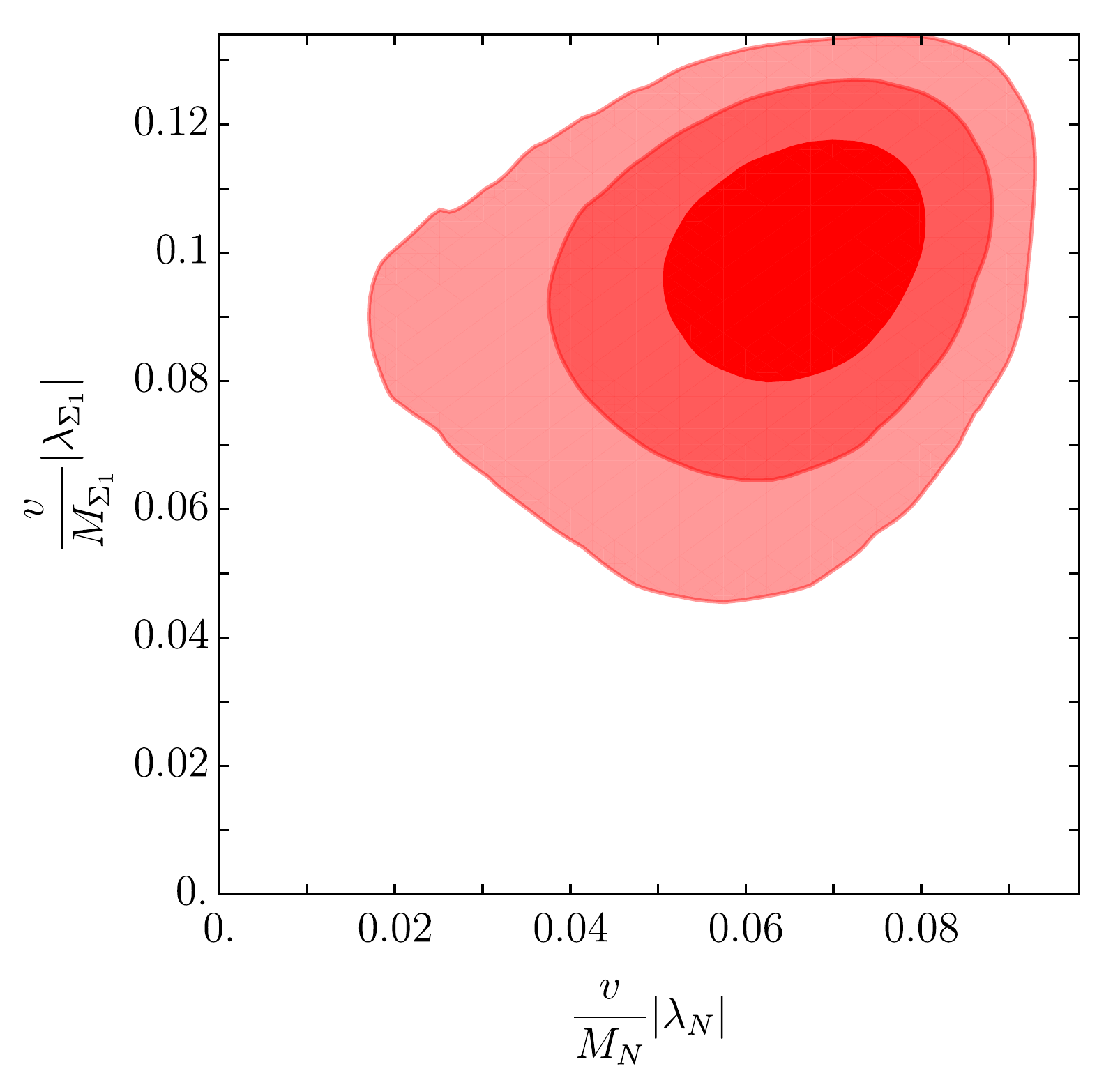}
\end{minipage}
\begin{minipage}[t]{.5\textwidth}
\phantom{ghost}\vspace{-3mm}
\includegraphics[width=.95\textwidth]{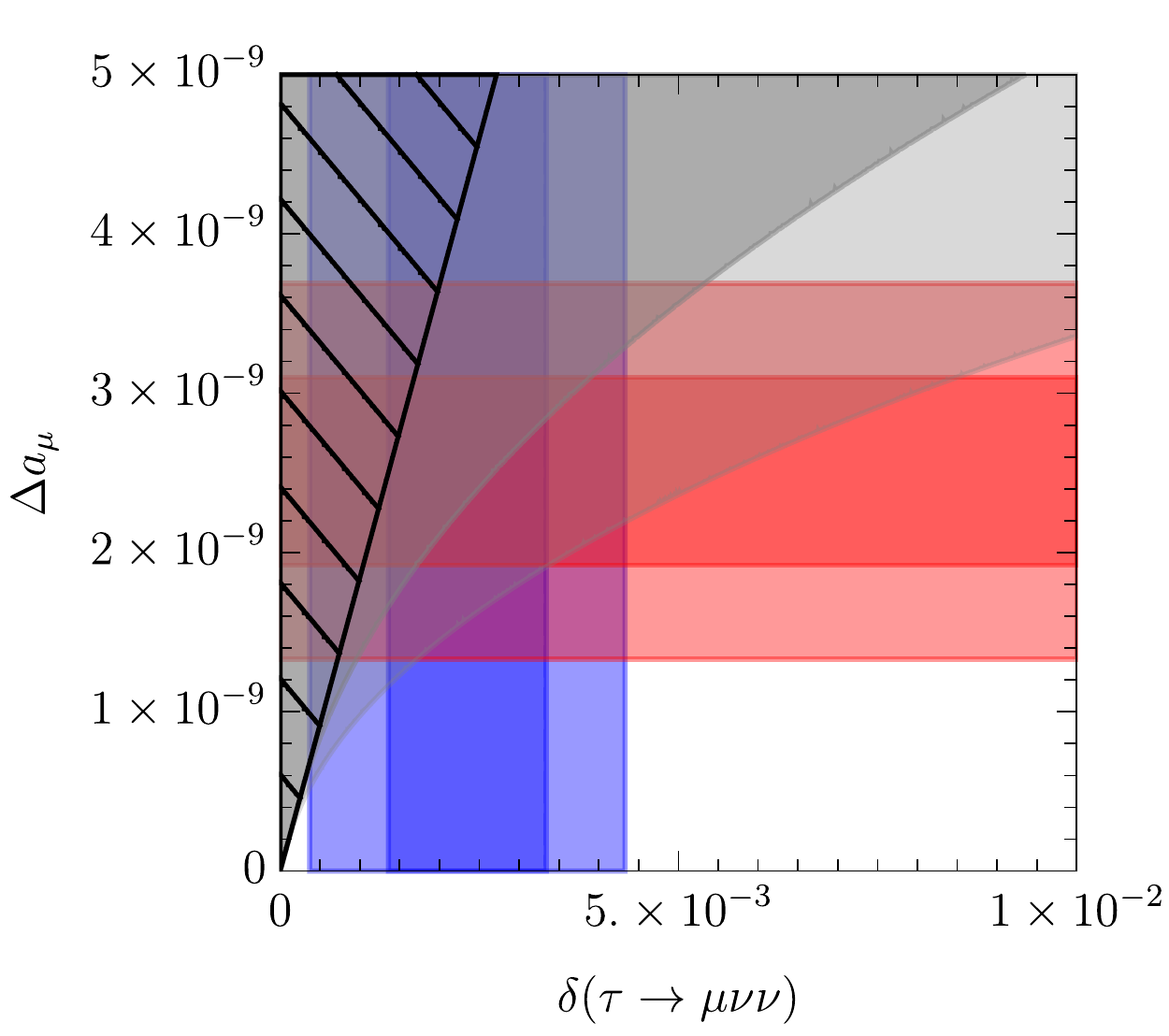}
\end{minipage}\\
\centering
\includegraphics[width=.45\textwidth]{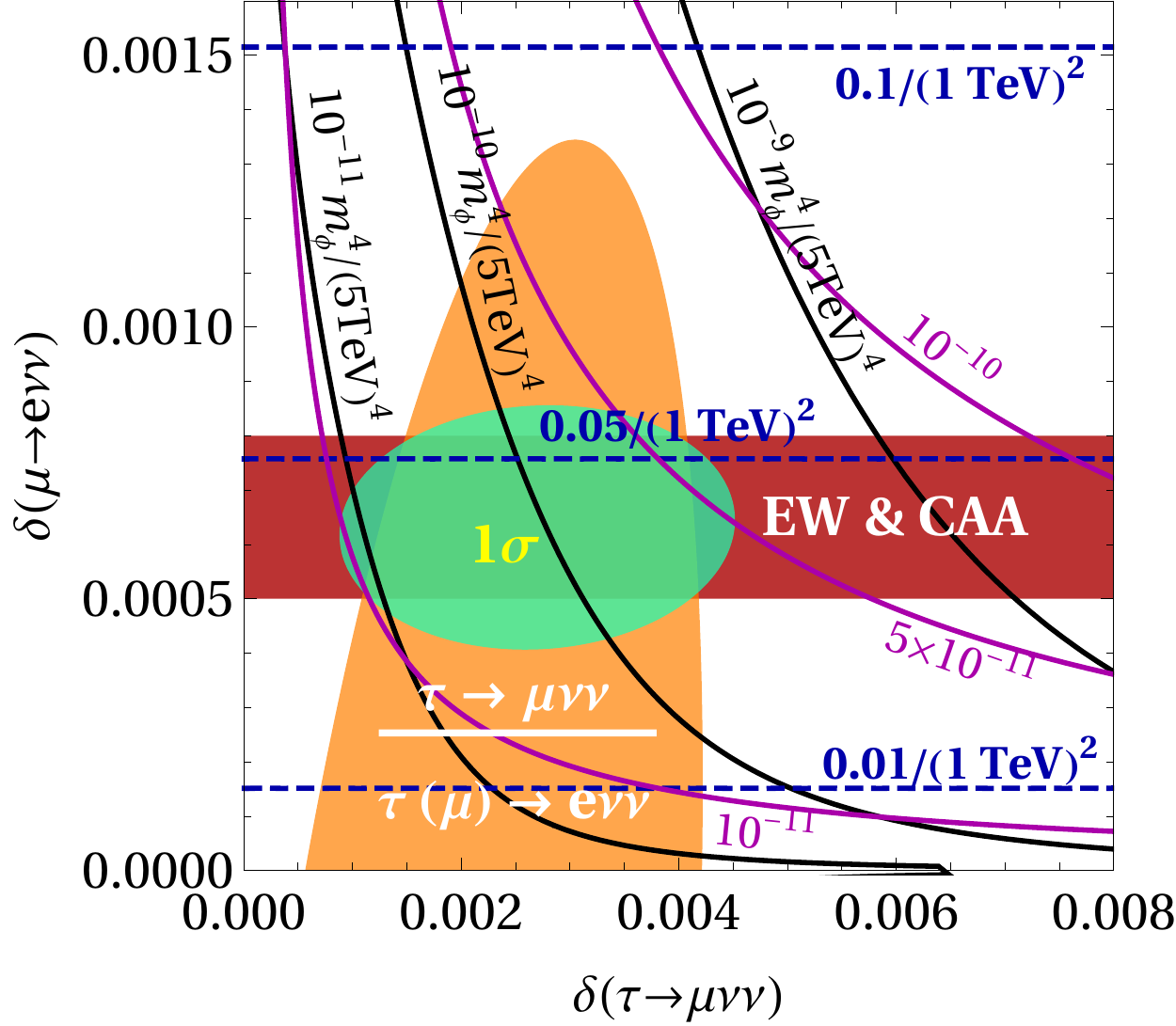}
\caption{
\textbf{top left:} Global electroweak fit of the VLL $N$ with mass $M_N$ and a coupling $\lambda_N$ to electrons and the VLL $\Sigma_1$ with mass $M_{\Sigma_1}$ and a coupling $\lambda_{\Sigma_1}$ to muons. The regions of 68\%, 95\% and 99.7\% C.L. are indicated in red. $v$ is the vev of the Higgs field.
\textbf{top right:} Correlations between the anomalous magnetic moment of the muon (allowed 1 and 2$\sigma$ regions shown in red) and $\tau\to \mu\nu\bar\nu$ (blue) in the simplified scenario of a $Z'$ boson with a mass of 1 TeV coupling with vectorial flavour off-diagonal couplings to muons and taus only. The regions excluded to 1 and 2$\sigma$ by electroweak data are shown in grey. The hatched region cannot be reached in this minimal setup. 
\textbf{bottom:} Regions preferred at the level of 1$\sigma$ by electroweak data and the Cabibbo Angle Anomaly (red) and the ratios given in Eq.~(\ref{eq:LFUratios}) (orange), as well as the combined 1$\sigma$ preferred region (green) shown in the plane of $\delta(\tau\to\mu\nu\bar\nu)$ and $\delta(\mu\to e\nu\bar\nu)$, where
 $\delta(\ell\to \ell'\nu\bar \nu)\equiv \mathcal{A}_{NP}(\ell\to \ell'\nu\bar \nu)/\mathcal{A}_{SM}(\ell\to \ell'\nu\bar \nu)$ is the new physics $\ell\to \ell'\nu\bar \nu$ amplitude, normalised to the same amplitude in the SM. Predictions for $\mathrm{Br}(\tau\to e\gamma)$ are shown in purple, predictions for $\mathrm{Br}(\tau\to e\mu\mu)$ in black and predictions for $|\lambda_{e\mu}|^2$, where $\lambda_{e\mu}$ is the coupling of the singly charged scalar singlet to the electron and muon generations, are indicated by blue dashed lines. $|\lambda_{e\mu}|^2$ can be probed by mono-photon searches at future $e^+e^-$ colliders. 
\label{fig:results}
}
\end{figure}

\paragraph{Vectorlike Leptons (VLLs)}
are hypothetical fermions that are neutral under $SU(3)_c$ and whose left- and right handed components have equal quantum numbers, making them anomaly-free extensions of the SM. 
After electroweak symmetry breaking, they can mix with the SM leptons, thus modifying the $W$ and $Z$ gauge boson couplings to the SM leptons (see $\delta (W\ell\nu)$ in Figures~\ref{fig:CAAleptophilic} and \ref{fig:LFUVleptophilic}). Single representations of VLLs do not give a good fit to data and are strongly constrained by flavour violating processes, if they couple to more than one generation of SM leptons. An excellent fit to data is achieved by allowing a VLL triplet of hypercharge $Y=-1$ to couple only to muons and a neutral singlet VLL $N$ to couple only to electrons. This scenario features enhanced $W\mu\nu$ couplings, that can address the Cabibbo Angle Anomaly, as well as decreased $Z\nu_e\nu_e$ couplings, which counterbalance the enhanced muon neutrino contributions to $Z\to\nu\nu$, while evading the very stringent constraints from $\mu\to e$ transitions. As is shown in the top right in Figure~\ref{fig:results}, it is preferred over the SM by over 4$\sigma$~\cite{Crivellin:2020ebi,DeBlas:2019ehy}.

\paragraph{The Singly Charged Scalar Singlet}\hspace{-3mm}\cite{Zee:1980ai}, $\phi^+$, can only interact with matter via
\begin{equation}
\mathcal{L}_{int} = - \frac{\lambda_{ij}}{2}\, \bar{L}^c_{a,i}\, \varepsilon_{ab}\, L_{b,j} \, \phi^+ + {\rm h.c.}\,.
\end{equation}
where $a,b$ are $SU(2)_L$ indices, $\varepsilon_{ab}$ is the 2-dimensional Levi-Civita tensor, $c$ stands for charge conjugation and $i,j$ are flavour
indices. The presence of the antisymmetric Levi-Civita tensor in this interaction term implies that the couplings $\lambda_{ij}$ can be chosen to be antisymmetric in flavour without loss of generality, making the interactions of the singly charged singlet scalar with leptons necessarily lepton flavour violating. If the three non-zero couplings $\lambda_{e\mu}$, $\lambda_{e\tau}$ and $\lambda_{\mu\tau}$ are not all equal, the singly charged scalar singlet can also induce lepton flavour universality violation.

Since a non-zero $\lambda_{e\mu}$ coupling is required for explaining the Cabibbo Angle Anomaly, and a non-zero $\lambda_{23}$ coupling is needed to generate an effect in $\tau\to \mu\bar\nu\nu$, we choose to set $\lambda_{e\tau}$ to zero in order to evade the bounds from $\mu\to e$ transitions.
Combining the data given in Eq.~(\ref{eq:LFUratios}) with the region preferred by electroweak data and the Cabibbo Angle Anomaly~\cite{DeBlas:2019ehy}, and superposing predictions for $\tau\to e\gamma$, $\tau\to e\mu\mu$ and $|\lambda_{e\mu}|^2$, we find the situation shown at the bottom in Figure~\ref{fig:results}~\cite{Crivellin:2020klg}.

\paragraph{Leptophilic $Z'$ Bosons}
are less constrained~\cite{ALEPH:2013dgf} than $Z'$ bosons, which also couple to quarks.
Following an agnostic approach, we set the mass of the leptophilic $Z'$ to 1 TeV and allow both for flavour diagonal and flavour off-diagonal couplings. While in generic scenarios our electroweak fits~\cite{DeBlas:2019ehy,Buras:2021btx} only displayed weak correlations between the $Z'$ couplings~\cite{Buras:2021btx}, simpler scenarios, such as a $Z'$ boson coupling only flavour off-diagonally to muons and taus, can, via the left-handed coupling, lead to new physics in $\tau\to \mu\bar\nu\nu$ and, via the product of the left- and the right-handed couplings, generate an $m_\tau /m_\mu$-enhanced effect in the anomalous magnetic moment of the muon, $\Delta a_\mu$. A simultaneous explanation of both the LFU ratios in Eq.~(\ref{eq:LFUratios}) and the discrepancy between theory and experiment in the anomalous magnetic moment of the muon, $\Delta a_\mu$~\cite{Muong-2:2021ojo,Aoyama:2020ynm} can be achieved without violating bounds from electroweak data (see top right in Figure~\ref{fig:results}).

\section*{Acknowledgments}
I would like to thank the organisers of the 32$^{nd}$ Rencontres de Blois for an inspiring conference in a beautiful chateau, and Bolek Pietrzyk, as well as my supervisor Andreas Crivellin for giving me the opportunity to give this talk.

\end{document}